\documentclass[12pt]{article}
\usepackage[numbers]{natbib}
\usepackage[dvips]{graphicx}
\usepackage{subcaption}
\usepackage[margin=1in,top=0.8in]{geometry}
\usepackage{hhline,amssymb,epsfig,amsmath,array,amsthm,color,setspace,titlesec,hyperref,lipsum,enumitem}
\usepackage[toc]{appendix}
\usepackage[T1]{fontenc}
\usepackage[utf8]{inputenc}
\usepackage{authblk}
\usepackage{changes}
\usepackage{algorithm, algorithmic}
\usepackage{mathtools}
\usepackage{tabulary}
\usepackage{diagbox}
\usepackage{bbm}
\usepackage{multirow}
\usepackage{natbib}
\newtheorem{theorem}{Theorem}
\newtheorem{corollary}{Corollary}
\newtheorem{assumption}{Assumption}
\newtheorem{remark}{Remark}

\DeclareMathOperator*{\argmax}{arg\,max}
\DeclareMathOperator*{\argmin}{arg\,min}

\begin{document}

\title{Crowdsourcing Utilizing Subgroup Structure of Latent Factor Modeling}
\author{Qi Xu, Yubai Yuan, Junhui Wang and Annie Qu\footnote{Qi Xu is Ph.D. student, Department of Statistics, University of California Irvine, Irvine, CA 92697 (E-mail: qxu6@uci.edu). Yubai Yuan is Assistant Professor, Department of Statistics, Penn State University, University Park, PA 16802 (E-mail: yvy5509@psu.edu). Junhui Wang is Professor, Department of Statistics, Chinese University of Hong Kong, Shatin, Hong Kong(E-mail: junhuiwang@cuhk.edu.hk). Annie Qu is Chancellor's Professor, Department of Statistics, University of California Irvine, Irvine, CA 92697 (E-mail: aq2@uci.edu).}}
\date{}
\maketitle

\begin{abstract}
\singlespacing

Crowdsourcing has emerged as an alternative solution for collecting large scale labels. However, the majority of recruited workers are not domain experts, so their contributed labels could be noisy. In this paper, we propose a two-stage model to predict the true labels for multicategory classification tasks in crowdsourcing. In the first stage, we fit the observed labels with a latent factor model and incorporate subgroup structures for both tasks and workers through a multi-centroid grouping penalty. Group-specific rotations are introduced to align workers with different task categories to solve multicategory crowdsourcing tasks. In the second stage, we propose a concordance-based approach to identify high-quality worker subgroups who are relied upon to assign labels to tasks. In theory, we show the estimation consistency of the latent factors and the prediction consistency of the proposed method. The simulation studies show that the proposed method outperforms the existing competitive methods, assuming the subgroup structures within tasks and workers. We also demonstrate the application of the proposed method to real world problems and show its superiority.

\medskip
\medskip
\noindent Key words: Bi-clustering; Concordance-based Approach; Label aggregation; Low rank approximation.

\end{abstract}

\newpage
\section{Introduction}

In recent years, there has been a growing demand for large-scale data with high-quality labels, which is essential for training high-performance machine learning models. However, relying completely on experts to label large-scale data could be very costly and time-consuming. As a popular substitution, crowdsourcing aims to predict true labels using a large number of crowd labels contributed by crowd workers who are assigned tasks; for example, in image classification tasks, crowd workers are required to classify tasks into categories and specify detailed labels. In the past two decades, benefiting from the development of online platforms like MTurk\footnote{https://www.mturk.com}, crowdsourcing has been applied to many label-intensive fields, for example, computer vision \citep{deng2009imagenet, kovashka2016crowdsourcing, su2012crowdsourcing}, natural language processing \citep{borromeo2015automatic, maclean2013identifying, snow2008cheap}, and medical diagnostics \citep{foncubierta2012ground, li2017reliable, mitry2015crowdsourcing}. The main challenge of crowdsourcing is that crowd labels might be noisy due to varying difficulty levels of tasks and diverse experience of workers on specific tasks. For example, high-quality workers can label tasks more accurately with high probability, while low-quality workers' performances are more like random guessing in labeling tasks. Therefore, the essence of effective crowdsourcing requires incorporating heterogeneity among tasks and workers to improve the label prediction accuracy.

One naive approach of crowdsourcing is majority voting \citep{aydin2014crowdsourcing, berend2014consistency, nitzan1982optimal}, which designates the labels voted by a majority of workers as the predicted labels. However, majority voting is suboptimal when the proportion of low-quality workers is relatively large. To incorporate the heterogeneity among workers, confusion matrix-based models have been developed recently; such as maximum likelihood via EM algorithm \citep{dawid1979maximum, zhang2016spectral}, bayesian classifier combination \citep{kim2012bayesian, venanzi2014community}, and pairwise co-occurrence factorization \citep{ibrahim2019crowdsourcing}. However, these approaches neglect the heterogeneity among tasks and assume that the probabilities of correct labeling are the same for tasks in the same category. To accommodate the heterogeneity of task difficulties and characteristics, latent factor models have been proposed; for example, the generative model of labels, abilities, and difficulties (GLAD) \citep{whitehill2009whose} and the multi-dimensional wisdom model \citep{welinder2010multidimensional}. That is, every task or worker is represented by a latent factor that encodes their characteristics, such as, task difficulty and worker ability \citep{whitehill2009whose}. Although existing latent factor models allow for heterogeneity among tasks and workers simultaneously, they are only applicable to binary tasks and not to multi-category tasks.

Another key challenge in crowdsourcing is label specification. A method that can correctly cluster tasks into categories might still fail to correctly specify labels for each category. For example, for confusion matrix-based models \citep{dawid1979maximum, zhang2016spectral}, the likelihood function remains unchanged even if the labels of categories are randomly shuffled. Thus, assumptions are generally required for category-label specification. One popular assumption adopted by confusion matrix-based models \citep{dawid1979maximum, zhang2016spectral} is to assume that labeling by majority voting is correct. Alternatively, \citet{ghosh2011moderates} assume that a high-quality worker is known who is capable of labeling categories better than random guessing. However, these assumptions are relatively restrictive and might not be satisfied when low-quality workers are prevalent, or a high-quality worker might not be known. This motivates us to develop a new framework for recover the category-label specification, so that the proposed method can be more robust against low-quality workers without sacrificing label prediction accuracy.

In this paper, we propose a novel crowdsourcing framework unifying binary and multicategory label prediction. Specifically, we embed tasks and workers into the same latent space, which is able to model crowd labels as a categorical distribution parameterized by the concordance between corresponding task and worker latent representations. One major innovation of the proposed method is that it incorporates heterogeneity among tasks and workers with latent factor modeling, while encouraging similar tasks and workers to cluster together. A similar concept has been utilized in the confusion matrix-based method \citep{venanzi2014community}, but the subgroup structures of tasks are ignored due to the nature of confusion matrix-based modeling. To capture subgroup structures of tasks and workers simultaneously, we introduce a multi-centroid grouping penalty, which can adaptively capture subgroup structures among tasks and workers and push latent factors of tasks or workers from the same subgroups together. To accommodate the multicategory labels, we introduce a set of group-specific rotation matrices to align latent factors of workers with latent factors of tasks from different categories. For the recovery of category-label specification, we identify high-quality workers through a concordance-based rule, which is capable of assigning labels accurately and robustly.

Compared with existing crowdsourcing methods, the proposed method has two main advantages. First, the proposed method improves label prediction accuracy through utilizing the subgroup structures in the latent space. The subgrouping penalty aims to capture intrinsic subgroup structures, and also differentiate different task categories and worker groups within the latent space. As a byproduct, the subgrouping penalty imposes constraints on latent factors so that the model complexity can be controlled to alleviate possible model overfitting to noisy crowd labels. In addition, the group-specific rotation matrices capture the common opinions regarding different task categories for workers from the same group, which utilize the subgroup structures of workers to induce a parsimonious model structure. From the perspective of statistical efficiency, the subgrouping penalty and group-specific rotation matrices improve estimation efficiency through borrowing information from the same subgroup.

Second, the proposed label specification is still valid when the proportion of low-quality workers is dominant or the high-quality worker is unknown. This is because the proposed method is able to simultaneously identify the high-quality worker subgroup and recover correct category-label specification, based on the assumption that high-quality workers have more certainty to label correctly than others. The labeling uncertainty is characterized by the concordance between tasks and worker subgroups in the latent space, regardless of the size of a worker subgroup and the subgroup membership of an individual worker. Thus, the proposed method is especially effective for settings where a small group of experts is mixed with a large number of low-quality workers.

To implement the proposed method, we propose a scalable algorithm to solve the non-convex optimization problem. To tune the hyper-parameters in the model, we propose a BIC-type criterion based on the labels given by the selected high-quality worker subgroups, which are assumed to be a surrogate of true labels. The simulation studies illustrate that the proposed method outperforms the existing competitive methods especially when low-quality workers are dominant among all workers. Theoretically, we show that model parameters are consistent under the hellinger distance and further prove the consistency of the label prediction.

The rest of this paper is organized as follows. Section 2 provides the background of latent factor-based crowdsourcing models and introduces the notations used throughout the paper. Section 3 presents the proposed framework for multicategory crowdsourcing, and provides the corresponding label specification procedures. Section 4 illustrates the scalable optimization algorithm and implementation details for the proposed method. Section 5 establishes the theoretical properties for the proposed method. In Section 6, we perform numerical comparisons between the proposed method and existing methods under simulation settings. In Section 7, we investigate the performance of the proposed method on multiple crowdsourcing benchmark datasets. Section 8 provides some discussion of the proposed method and concluding remarks.

\section{Background and Notations} 

In this section, we provide the background of the crowdsourcing approach based on latent factor modeling, and introduce the notations used throughout the paper. We define $\mathbf{R} = \{r_{ij}\}_{m\times n}$ as a crowd label matrix, where $r_{ij}$ is the label for the $i$th task given by the $j$th worker ($i = 1,...,m; j = 1, ..., n$). We assume $r_{ij} \in \{0, 1\}$ for binary crowdsourcing and $r_{ij} \in \{0, 1, ..., C-1\}$ for multicategory crowdsourcing with $C$ categories. In practice, workers typically do not label all tasks in that only a subset of $\mathbf{R}$ is observed, and we denote the subset as $\Omega = \{(i, j): r_{ij} \text{ is observed}\}$. We denote $\mathbf{Z} = (Z_1, Z_2, ..., Z_m)'$ as the true labels for tasks, and if $Z_i = c$, we name the $i$th task as label-$c$ task. The goal of crowdsourcing is to predict $Z_i$'s via aggregating observed $r_{ij}$'s.

A seminal crowdsourcing approach, the generative model of label, ability and difficulty (GLAD) \citep{whitehill2009whose}, is a logistic model based on task and worker features:
\begin{align}
    \label{eq: glad}
     \log \left\{ \frac{P(r_{ij} = Z_i)}{1 - P(r_{ij} = Z_i)}\right\} = \mu_i\nu_j,
 \end{align}
 where $\mu_i \in (0, \infty)$ is a parameter measuring the difficulty level of the $i$th task and a larger $\mu_i$ indicates less difficulty of the $i$th task, and $\nu_j \in (-\infty, \infty)$ is a parameter measuring the $j$th worker's ability with a higher value associated with greater ability to label tasks. The probability $P(r_{ij} = Z_i)$ converges to $1$ if $\mu_i\nu_j \rightarrow \infty$, that is, the chance of correct labeling is high with a high quality worker for an easy task. The GLAD model incorporates the task and worker heterogeneity simultaneously by introducing two latent variables, but it's limited for binary crowdsourcing, and one-dimensional approximation is restricted in real applications. 

\section{Methodology}

In this section, we introduce our proposed two-stage model for the multi-categorical crowdsourcing. The first stage is to fit the observed labels with a latent factor model, and the second stage is to identify high-quality workers and to make predictions. The binary crowdsourcing problem is a special case under our framework, and detailed discussions about binary crowdsourcing are provided in supplementary materials.

\subsection{Model Framework}

For a $C-$category crowdsourcing problem, we assume that $r_{ij}$'s follow a categorical distribution:
\begin{align}
    \label{dist_label}
    r_{ij} \sim categorical(p_{ij}^{(0)}, p_{ij}^{(1)}, ..., p_{ij}^{(C-1)}), \quad \sum_{c=0}^{C-1}p_{ij}^{(c)}=1, \quad (i,j)\in \Omega,
\end{align}
where $p_{ij}^{(c)}$ is the probability of $r_{ij}$ equal to $c$. Here we model the probabilities of the observed labels, rather than the probability of correctly labeling $\mathbbm{1}(r_{ij}=Z_{i})$ as in (\ref{eq: glad}). Our approach is more flexible for modeling multicategory labels.

To capture the heterogeneity within tasks and workers, a full parameterization model is introduced:
\begin{align}
   \label{eq: full parameterization model}
   \log\left\{\frac{p_{ij}^{(c)}}{p_{ij}^{(0)}}\right\} = \mathbf{a}_i'\mathbf{b}_{j, c} - \mathbf{a}_i'\mathbf{b}_{j, 0}, \quad  c = 1,..., C-1, \quad (i, j) \in \Omega,
\end{align}
where $\mathbf{a}_i \in \mathbb{R}^{k}$ is the latent factor for the $i$th task, representing the latent characteristics of this task. In the examples of image classification problems, $\mathbf{a}_{i}$'s might encode the object and texture information. The latent factor $\mathbf{b}_{j,c} \in \mathbb{R}^{k}$ represents the $j$th worker's opinion on the label-$c$ task. The inner product $\mathbf{a}_i'\mathbf{b}_{j,c}$ measures the concordance between the $i$th task and the label-$c$ task based on the $j$th worker's opinion. In contrast to the GLAD (\ref{eq: glad}), the true labels $Z_{i}$'s are implicitly embedded in $\mathbf{a}_{i}$'s, and tasks in the same category are likely to share similar latent features. Therefore, the latent factors of tasks are expected to be clustered, where we use $\mathbf{U} = \{U_{i}\}_{i=1}^{m}$ ($U_i \in \{0, 1, ..., C-1\}$) to denote the task memberships. Note that task membership $\mathbf{U}$ is a permutation of true labels $\mathbf{Z}$, and the correspondence between $\mathbf{U}$ and $\mathbf{Z}$ is recovered in the second stage.

The full parameterization model (\ref{eq: full parameterization model}) is similar to polytomous logistic regression \citep{engel1988polytomous} where each category $c$ corresponds to a set of category-specific parameters $\mathbf{b}_{j, c}$. However, the full parameterization model (\ref{eq: full parameterization model}) has several disadvantages when it is applied to crowdsourcing problems. First, it fails to capture the correlation among different workers, which are widely observed in real world applications \citep{cao2019max, li2019exploiting}; for example, workers having the same education background are likely to have similar opinions on specific tasks. Second, the model (\ref{eq: full parameterization model}) is over-parameterized where each worker is associated with $C$ latent factors, which leads to an inefficient estimation of the latent factors of workers since it can only utilize the information from the corresponding task category.

\begin{figure}
  \centering
    \includegraphics[width=0.7\textwidth]{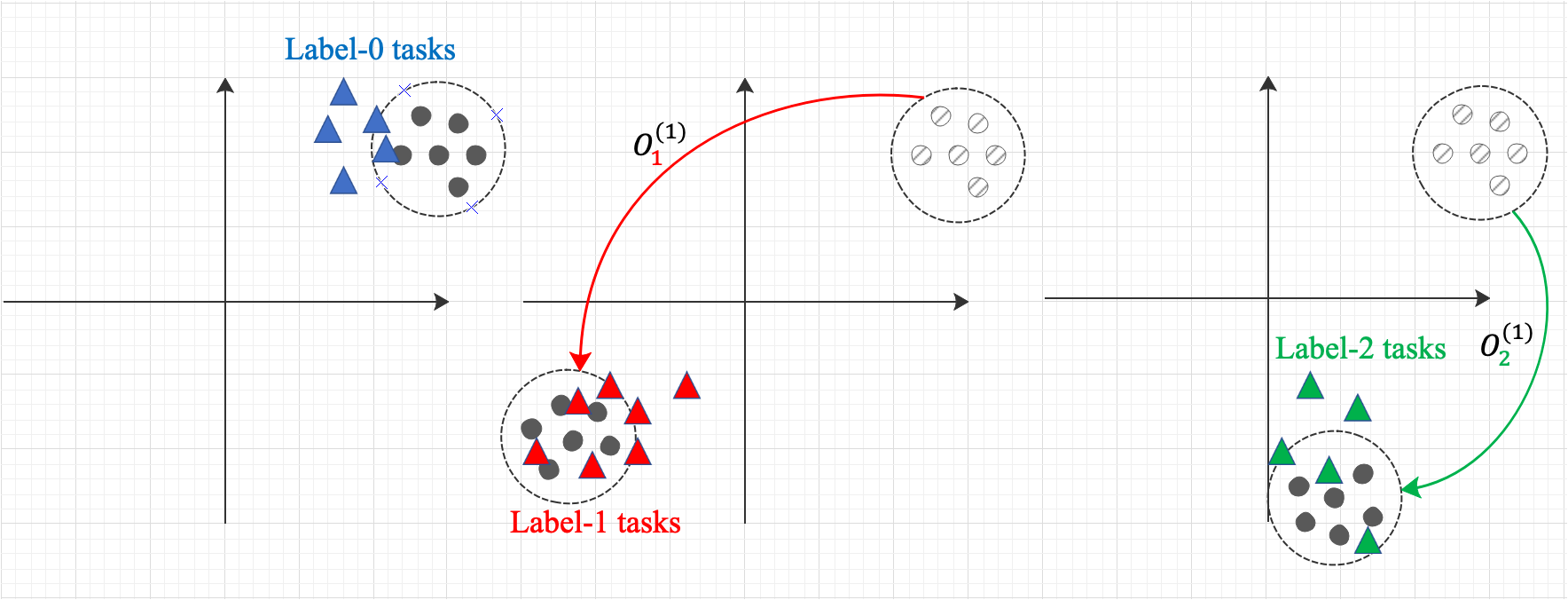}
  \caption{Illustration of the proposed multicategory model (\ref{eq: multi-categorical model}). Triangles in the three panels are latent factor of label-0, label-1 and label-2 tasks respectively. The dots in the left panel are latent factors of a group of workers, representing opinions on the label-$0$ tasks. Rotated dots in the middle and right panels represent the workers' opinions on the label-$1$ and label-$2$ tasks, respectively.}
  \label{fig: multi_model}
\end{figure}

To tackle the above problems, we propose a novel approach that leverages the subgroup structures within tasks and workers. Suppose we have $D$ worker groups, and the group membership of workers is denoted by $\mathbf{V} = \{V_{j}\}_{j=1}^{n}$ where $V_{j} \in \{1, 2, ..., D\}$. Workers from the same subgroup tend to have similar opinions over tasks, and within-group correlations among workers' opinions should be taken into consideration. For worker subgroup $d \in \{1, 2, ..., D\}$, we introduce a set of rotation matrices 
\begin{align}
    \label{eq: rotation_matrices}
    \mathbf{O}^{(d)} = \{\mathbf{O}^{(d)}_c \mid \mathbf{O}^{(d)^{'}}_{c}\mathbf{O}^{(d)}_{c} = \mathbf{I}_{k\times k}, \mathbf{O}^{(d)}_c \in \mathbb{R}^{k\times k}, c = 0, 1, ..., C-1\},
\end{align}
to align the latent factors of the $d$th worker subgroup with the latent factors of each task category. Specifically, for the $j$th worker belonging to the $d$th subgroup, its opinion on the label-$c$ tasks is modeled based on rotating its opinion on the reference category (label-$0$ tasks are treated as reference in default in this paper) via the rotation matrix $\mathbf{O}_c^{(d)}$:
{\footnotesize
\begin{align}
\label{eq: rotated_latent_factor}
   \mathbf{b}_{j, c} = \mathbf{O}^{(d)}_c\mathbf{b}_{j}, \quad c = 0, 1, ..., C-1, \quad d = V_{j},
\end{align}
}%
where we specify $\mathbf{O}^{(d)}_0 = \mathbf{I}_{k\times k}$ for an identifiability purpose such that $\mathbf{b}_j =  \mathbf{b}_{j, 0}$ represents the $j$th worker's opinion on the reference group. We denote all the rotation matrices as $\mathbf{O} = \{\mathbf{O}^{(1)}, \mathbf{O}^{(2)}, ..., \mathbf{O}^{(D)}\}$. Based on the proposed modeling for $\mathbf{b}_{j, c}$ in (\ref{eq: full parameterization model}), the multicategory crowd labels follow a logistic model:
{\footnotesize
\begin{equation}
   \label{eq: multi-categorical model}
   \theta_{ijc} = \log\left\{\frac{P(r_{ij} = c)}{P(r_{ij} = 0)}\right\} = \mathbf{a}_i'{\mathbf{O}}^{(d)}_c\mathbf{b}_j - \mathbf{a}_i'\mathbf{b}_j, \quad  c = 1,..., C-1, \quad d = V_j, \quad (i, j) \in \Omega.
\end{equation}
}%
For ease of notation, we denote $\mathbf{A} = (\mathbf{a}_1, \mathbf{a}_2, ..., \mathbf{a}_m)'$ as collections of latent factors for tasks. Similarly, we use $\mathbf{B} = (\mathbf{b}_1, \mathbf{b}_2, ..., \mathbf{b}_n)'$ as collections of latent factors for workers. Figure \ref{fig: multi_model} illustrates the proposed model (\ref{eq: multi-categorical model}) in that the rotation matrices $\mathbf{O}^{(1)}_1$ and $\mathbf{O}^{(1)}_2$ align the first subgroup workers' opinions on reference tasks with the latent factors of the label-1 and label-2 tasks, respectively.

The proposed method has the following advantages over the full parameterization model (\ref{eq: full parameterization model}). First, the proposed model is able to capture the underlying correlation among workers in the same subgroup via $\mathbf{O}_d^{(c)}$'s shared by worker subgroups. Incorporating these dependencies enables us to borrow information from observed crowd labels from workers in the same subgroup, to improve estimation accuracy and efficiency. Second, the proposed method can detect and enhance the subgroup structures among workers, which is an important merit for crowdsourcing. Specifically, the embedded subgrouping procedure will allow us to identify the high-quality worker subgroup for each task category, and further improve label prediction accuracy based solely on the high-quality worker subgroups. Third, the proposed method is more parsimonious than the full parameterization model in that the proposed method requires $(m+n)\times k + (C-1)\times k^2$ parameters while the full parameterization model requires $(m + nC)\times k$. The advantage is more substantial especially when $n \gg k$, which is common in practice.

However, fitting model (\ref{eq: multi-categorical model}) does not provide a direct prediction on true labels $\mathbf{Z}$ but recovering the subgroup memberships of tasks $\mathbf{U}$. Therefore, in the second stage, the proposed method aims to recover the correct correspondence between task memberships $\mathbf{U}$ and true labels $\mathbf{Z}$. For a $C$-category crowdsourcing problem, existing works \citep{dawid1979maximum, zhang2016spectral} assume most workers can label tasks correctly for all categories with a high probability, which is quite restrictive in practice. Instead, we relax the assumption which only requires a subset of workers to distinguish some categories well with high accuracy. Consequently, we propose to identify the high-quality worker subgroup for each task category and recover the correspondence based on labels from high-quality workers.

\begin{figure}
  \centering
  \includegraphics[width=0.7\textwidth]{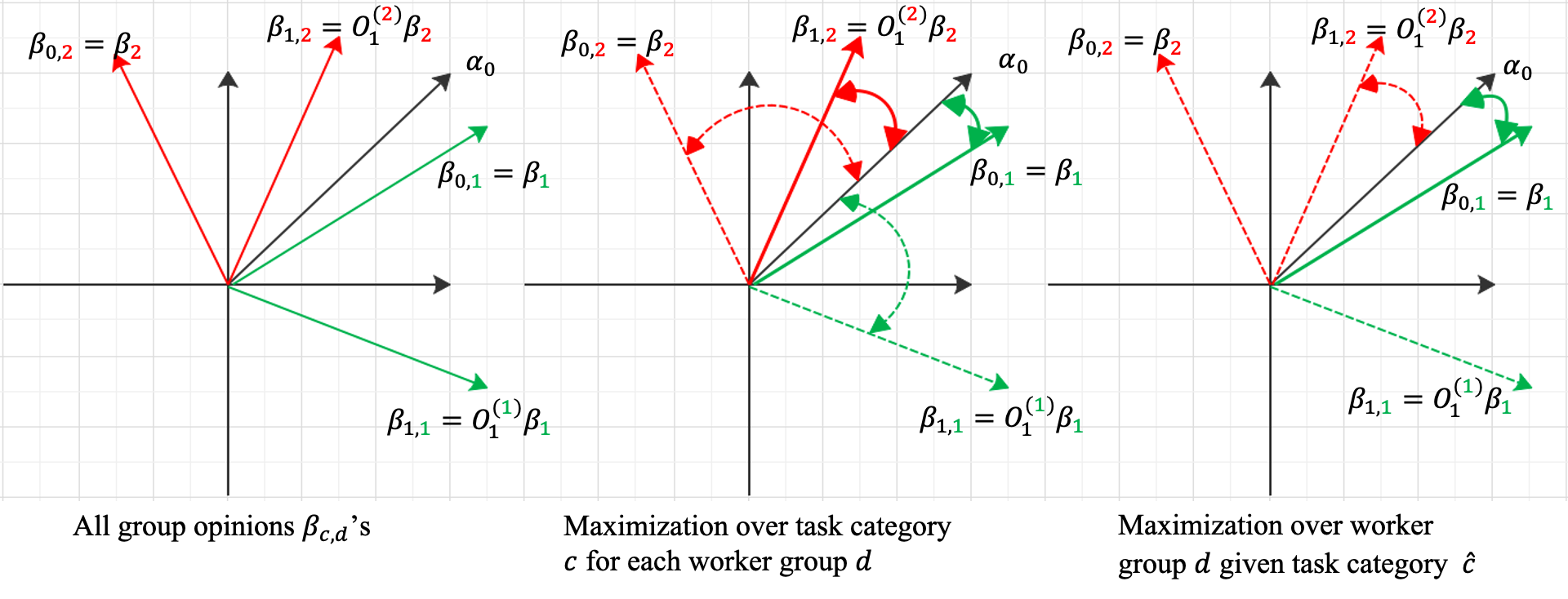}
  \caption{An example of identifying high-quality workers and assigning labels to $\alpha_0$. Given two worker groups, latent factor centroids $\beta_1, \mathbf{O}_{1}^{(1)}\beta_1$, $\beta_2, \mathbf{O}^{(2)}_1\beta_2$ represent the opinions of two worker groups on two task categories. Based on Assumption \ref{assumption 2} and the high-quality worker identification criterion, the centroid $\beta_1$ is the high-quality worker group for $\alpha_0$, and tasks in the $\alpha_0$ category are specified as label-$0$ tasks.}
  \label{fig: multi-assumption}
\end{figure}

For many domain-specific applications \citep{mitry2015crowdsourcing, foncubierta2012ground}, high-quality workers tend to have a greater certainty in providing correct labels than other workers. In the proposed latent factor model (\ref{eq: multi-categorical model}), the labeling certainty of a worker can be quantified as the concordance between the latent factors for the worker and the task. For example, a worker subgroup close to the centroid of a specific task in latent space has a greater labeling certainty than other worker subgroups. In the following, we denote $\boldsymbol\alpha_{I}, I=0, 1, ..., C-1$ as latent factor centroids of tasks, and $\boldsymbol\beta_{I, J}, I=0, 1, ..., C-1, J=1, 2, ..., D$ for worker subgroup $J$ with respect to each task category $I$. Figure \ref{fig: multi-assumption} illustrates the key concept of identifying the high-quality worker subgroup and assigning labels to the task category with centroid $\boldsymbol\alpha_0$. For the first worker subgroup, the centroid $\boldsymbol\beta_{0, 1}$ has the largest concordance with $\boldsymbol\alpha_0$, suggesting that the first worker subgroup classifies $\boldsymbol\alpha_0$ category tasks as label-$0$ tasks. For the second worker subgroup, the aligned centroid $\boldsymbol\beta_{1, 2}$ has the largest concordance with $\boldsymbol\alpha_0$, but it is smaller than the concordance between $\boldsymbol\beta_{0, 1}$ and $\boldsymbol\alpha_0$. Based on the above idea, we specify the unique label for each task subgroup, relying on the following assumption on the high-quality worker group:

\begin{assumption}
\label{assumption 2}
For each task category, a subgroup of workers is the high-quality worker group for this task category if the group members have stronger concordance with this category tasks than other groups. The label specified for this task category from the high-quality worker group is the true labels.
\end{assumption}

Based on Assumption \ref{assumption 2}, $\boldsymbol\beta_{0, 1}$ is the centroid of the high-quality worker subgroup for $\boldsymbol\alpha_0$, and $\boldsymbol\alpha_0$ is therefore specified as label-$0$ task by the worker group associated with $\boldsymbol\beta_{0, 1}$. In general, we can identify the high-quality worker subgroup and specify labels for the $i$th task by the following criterion based on Assumption \ref{assumption 2}:
{\footnotesize
\begin{align}
\label{eq: multi_criterion}
    (\hat{c}_i, \hat{d}_i) = \argmax_{\substack{c \in \{0, 1, ..., C-1\},\\ d \in \{1, 2, ..., D\}}} \boldsymbol\alpha_{U_i}'\boldsymbol\beta_{c, d},
\end{align}
}%
where $\hat{d}_i$ is the high-quality worker group for the $i$th task and $\hat{c}_i$ is the label assigned by the high-quality worker group $\hat{d}_i$. The maximization in (\ref{eq: multi_criterion}) can be done by profiling. For each fixed worker group $d = 1, 2, ..., D$, the maximization over $c$ corresponds to aligning the $d$th group workers' opinion to $\alpha_{U_i}$. The maximization over $d$ corresponds to identifying the high-quality worker group for the $i$th task.

The proposed method relies on the identification of high-quality workers and their category-label specification. Compared with the confusion matrix-based method \citep{dawid1979maximum, zhang2016spectral} adoption of the category-label specification based on majority voting, the proposed method is more robust against the proportion of low-quality workers. Furthermore, compared with \citet{ghosh2011moderates}'s requirement of known high-quality workers, the proposed method is a data-driven approach, as high-quality workers are directly identified from the data.

\subsection{Parameter Estimation}

In this subsection, we introduce how to obtain estimations of model parameters through labeling matrix $\mathbf{R}$. In addition, we impose a novel grouping penalty to pursue the subgroup structures in the latent space.

Based on the assumed categorical distribution of observed labels, the negative log-likelihood of observed labels are:
\begin{align}
    \mathcal{L}(\mathbf{A}, \mathbf{B}, \mathbf{O}) &= -\sum_{(i,j)\in\Omega}\sum_{c=0}^{C-1}\mathbbm{1}(r_{ij}=c)\log{\tilde{r}_{ij,c}} \notag \\
    &= -\sum_{(i,j)\in\Omega}\sum_{c=0}^{C-1}\mathbbm{1}(r_{ij}=c)\log{\frac{\exp(\mathbf{a}_{i}'\mathbf{O}_{c}^{d}\mathbf{b}_{j})}{\exp(\mathbf{a}_{i}'\mathbf{b}_{j}) + \sum_{c'=1}^{C-1}\exp(\mathbf{a}_{i}'\mathbf{O}_{c'}^{d}\mathbf{b}_{j})}}. \notag
\end{align}
Furthermore, we utilize a multi-centroid grouping penalty to pursue the subgroup structures for both tasks and workers. In particular, we define the multi-centroid grouping penalty as follows:
{\footnotesize
\begin{align}
  \label{eq: penalty}
  \mathcal{G}_{\lambda}(\mathbf{A}, \mathbf{B}, \mathbf{U}, \mathbf{V}) = \lambda \bigg\{\lVert \mathbf{A} -  \mathcal{U}\mathbf{A}\rVert_F^2 + \lVert \mathbf{B} - \mathcal{V}\mathbf{B}\rVert_F^2\bigg\},
\end{align}
}%
where $\lambda$ is a tuning parameter for penalization, matrices $\mathcal{U}$ and $\mathcal{V}$ are projection matrices associated with group memberships $\mathbf{U}$ and $\mathbf{V}$ by a one-to-one mapping \citep{hoaglin1978hat} to compute the subgroup centroids of tasks and workers, respectively. Specifically, the projection matrices $\mathcal{U}$ and $\mathcal{V}$ project each individual latent factor $\mathbf{a}_i$ and $\mathbf{b}_j$ onto its corresponding centroids, i.e., $\mathcal{U}\mathbf{a}_i = \boldsymbol\alpha_{U_i}$ and $\mathcal{V}\mathbf{b}_j = \boldsymbol\beta_{0, V_j}$. Therefore, the multi-centroid grouping penalty is equivalent to 
\begin{align}
    \mathcal{G}_{\lambda}(\mathbf{A}, \mathbf{B}, \mathbf{U}, \mathbf{V}) = \lambda\bigg\{\sum_{I=0}^{C-1}\sum_{\{i: U_i = I\}}\lVert \mathbf{a}_i - \boldsymbol\alpha_I \rVert^2 + \sum_{J=1}^{D}\sum_{\{j: V_j = J\}}\lVert \mathbf{b}_j - \boldsymbol\beta_{0, J} \rVert^2\bigg\}. \notag
\end{align}
This type of penalization term was first proposed in \citep{tang2020individualized} in the longitudinal setting, and has been applied for directed network community detection \citep{zhang2021directed}. In addition, other penalty functions such as fused lasso \citep{tibshirani2005sparsity} and OSCAR \citep{bondell2008simultaneous}, can be used for subgrouping latent factors; however, they could suffer from computation burden and estimation bias. Thus, we choose to use the multi-centroid grouping penalty in our proposed method.

The multi-centroid grouping penalty has the following advantages. First, it pursues subgroup structures in the latent space and reduces estimation bias and computation cost. Traditional penalties for subgrouping pursuit, for example, the fused lasso \citep{tibshirani2005sparsity} and OSCAR \citep{bondell2008simultaneous}, suffer from estimation bias and computational cost due to the pairwise comparison. In contrast, the multi-centroid grouping penalty reduces computation cost from $O(m^2 + n^2)$ to $O(m + n)$. Meanwhile, rather than pulling latent factors together even from different subgroups, the multi-centroid grouping penalty separates dissimilar latent factors to achieve subgrouping. Furthermore, it reduces estimation bias by introducing multiple shrinkage directions. Compared with the general $L_2$ penalty, the multi-centroid grouping penalty prompts the centroids in a data-adaptive way, which makes the proposed method incorporate observed data more flexibly than a standard approach to a fixed point.

By incorporating the multi-centroid grouping penalty, we can estimate latent factors $\mathbf{A}$, $\mathbf{B}$ and group memberships $\mathbf{U}, \mathbf{V}$ jointly through minimizing the negative penalized log-likelihood:
\begin{align}
\label{eq: joint_loss}
\footnotesize
  (\hat{\mathbf{A}}, \hat{\mathbf{B}}, \hat{\mathbf{O}}, \hat{\mathbf{U}}, \hat{\mathbf{V}}) & = \argmin_{\mathbf{A}, \mathbf{B}, \mathbf{O}, \mathbf{U}, \mathbf{V}} \mathcal{L}(\mathbf{A}, \mathbf{B}, \mathbf{O}) + \mathcal{G}_{\lambda}(\mathbf{A}, \mathbf{B}, \mathbf{U}, \mathbf{V}).
\end{align}
After estimating the model parameters, we can identify the high-quality worker subgroup in the second stage for each task category based on Assumption \ref{assumption 2}, and predict true labels via (\ref{eq: multi_criterion}). The detailed algorithm for minimizing (\ref{eq: joint_loss}) and implementations are provided in Section \ref{sec: algorithm}.
    
\section{Algorithm and Implementation}

\label{sec: algorithm}
    In this section, we introduce scalable algorithms to estimate the latent factors, rotation matrices and group memberships via alternatively minimizing loss function (\ref{eq: joint_loss}) for the multi-categorical case. In addition, we provide a new criterion to tune hyper-parameters when the true label information is not available.
    
    We first introduce the algorithm for solving (\ref{eq: joint_loss}). The optimization of (\ref{eq: joint_loss}) is rather challenging because of its non-convexity and orthogonal constraints on rotation matrices. In the following, we propose an alternating minimization algorithm to minimize the joint loss function in (\ref{eq: joint_loss}). Specifically, at the $(t+1)$th iteration, we update $(\mathbf{\hat{A}}, \mathbf{\hat{B}}, \hat{\mathbf{O}})$ sequentially given the estimation $(\hat{\mathcal{U}}^{(t)}, \hat{\mathcal{V}}^{(t)})$ as follows:
    \begin{equation}
        \begin{aligned}
        \label{al: update_lfm}
        \hat{\mathbf{A}}^{(t+1)} &= \arg\min_{\mathbf{A}}\mathcal{L}(\mathbf{A}, \hat{\mathbf{B}}^{(t)}, \hat{\mathbf{O}}^{(t)}) + \lambda\lVert \mathbf{A} - \hat{\mathcal{U}}^{(t)}\mathbf{A}\rVert_F^2,\\
        \hat{\mathbf{B}}^{(t+1)} &= \arg\min_{\mathbf{B}}\mathcal{L}(\hat{\mathbf{A}}^{(t+1)}, \mathbf{B}, \hat{\mathbf{O}}^{(t)}) + \lambda\lVert \mathbf{B} - \hat{\mathcal{V}}^{(t)}\mathbf{B}\rVert_F^2,\\
        \hat{\mathbf{O}}^{(t+1)} &= \arg\min_{\mathbf{O}}\mathcal{L}(\hat{\mathbf{A}}^{(t+1)}, \hat{\mathbf{B}}^{(t+1)}, \mathbf{O}),
        \end{aligned}
    \end{equation}
    where the minimization regarding $\mathbf{O}$ is realized via performing Cayley transformation \citep{wen2013feasible} which guarantees the orthogonal constraints are held in the iterations.
    Specifically, we introduce $\mathbf{G} = \{\mathbf{G}^{(1)}_{0}, \mathbf{G}^{(1)}_{1},$ $...,\mathbf{G}^{(1)}_{C-1}, ..., \mathbf{G}^{(D)}_{C-1}\}$ as gradients of $\mathbf{L}$ with respect to $\mathbf{O}$, where $\mathbf{G}^{(d)}_c = \frac{\partial \mathbf{L}}{\partial \mathbf{O}_c^{(d)}}$. We then define a set of matrices $\mathbf{S} = \{\mathbf{S}^{(1)}_0, \mathbf{S}^{(1)}_1, ..., \mathbf{S}^{(1)}_{C-1}, ..., \mathbf{S}^{(D)}_{C-1}\}$, where $\mathbf{S}^{(d)}_c = \mathbf{G}^{(d)}_c{\mathbf{O}^{(d)}_c}' - \mathbf{O}^{(d)}_c{\mathbf{G}^{(d)}_c}'$. Then $\mathbf{O}^{(d)}_c$'s are updated following the Cayley transformation \citep{wen2013feasible}:
    \begin{align}
        \label{al: cayley}
        \mathbf{O}^{(d)}_c \leftarrow (\mathbf{I} + \frac{\eta}{2}\mathbf{S}_c^{(d)})^{-1}(\mathbf{I} - \frac{\eta}{2}\mathbf{S}_c^{(d)})\mathbf{O}^{(d)}_c, \quad d=1,...,D, \quad c = 1, ..., C-1,
    \end{align}
    where $\eta$ is a positive learning rate. We update $\mathbf{O}_c^{(d)}$'s iteratively by (\ref{al: cayley}) until the algorithm converges. It is noticeable that $\mathbf{O}^{(d)}_c$'s remain orthogonal at each iteration for any positive value of $\eta$, and we set $\eta=0.1$ empirically which leads to a fast convergence while maintaining a low training error. In addition, the update of latent factors $\mathbf{a}_{i}$'s and $\mathbf{b}_{j}$'s can be paralleled to speed up the computation. Furthermore, the updates of projection matrices $\mathcal{U}$ and $\mathcal{V}$ are equivalent to cluster task and worker latent factors into $C$ and $D$ subgroups, respectively. Therefore, we update $\mathcal{U}$ and $\mathcal{V}$ with K-Means algorithm. The proposed algorithm for the multi-category crowdsourcing is outlined in Algorithm \ref{al2}.
    
    \begin{algorithm}[H]
        \caption{Alternating minimization algorithm}
        \label{al2}
    	\begin{algorithmic}
    	\STATE\textbf{Initialization}: Initialize $\mathbf{A}^{(0)}, \mathbf{B}^{(0)}, \mathbf{O}^{(0)}, \mathcal{U}^{(0)}, \mathcal{V}^{(0)}$; set $\lambda, k, D$, $t=0$, learning rate $\eta$ and stopping tolerance $\epsilon$.\\
    	\STATE\hspace{\algorithmicindent}\textbf{Step 1.} At the $(t+1)$th iteration, given $\hat{\mathbf{O}}^{(t)}, \hat{\mathcal{U}}^{(t)}, \hat{\mathcal{V}}^{(t)}$ from the $t$th iteration, update latent factors $(\mathbf{\hat{A}}^{(t+1)}, \mathbf{\hat{B}}^{t+1})$ via (\ref{al: update_lfm}).\\
    	\STATE\hspace{\algorithmicindent}\textbf{Step 2.} At the $(t+1)$th iteration, given $\hat{\mathbf{A}}^{(t+1)}, \hat{\mathbf{B}}^{(t+1)}$ from Step 1, update orthogonal matrices $\hat{\mathbf{O}}^{(t+1)}$ via (\ref{al: cayley}).\\
    	\STATE\hspace{\algorithmicindent}\textbf{Step 3.} At the $(t+1)$th iteration, given $\hat{\mathbf{A}}^{(t+1)}, \hat{\mathbf{B}}^{(t+1)}, \hat{\mathbf{O}}^{(t+1)}$ from Step 1 and Step 2, update projection matrices $({\hat{\mathcal{U}}}^{(t+1)}, {\hat{\mathcal{V}}}^{(t+1)})$ via K-Means algorithm.\\
        \STATE\hspace{\algorithmicindent}\textbf{Step 4.} Iterates Steps 1-3 until $|\{\mathcal{L}(\mathbf{\hat{A}}^{(t+1)}, \mathbf{\hat{B}}^{(t+1)}, \hat{\mathbf{O}}^{(t+1)}) + \mathcal{G}_{\lambda}(\mathbf{\hat{A}}^{(t+1)}, \mathbf{\hat{B}}^{(t+1)},{\hat{\mathcal{U}}}^{(t+1)},$ ${\hat{\mathcal{V}}}^{(t+1)})\} - \{\mathcal{L}(\mathbf{\hat{A}}^{(t)}, \mathbf{\hat{B}}^{(t)}, \hat{\mathbf{O}}^{(t)}) + \mathcal{G}_{\lambda}(\mathbf{\hat{A}}^{(t)}, \mathbf{\hat{B}}^{(t)}, {\hat{\mathcal{U}}}^{(t)}, {\hat{\mathcal{V}}}^{(t)})\}| \le \epsilon.$
    	\end{algorithmic}
    \end{algorithm}
    
    Since the loss function in (\ref{eq: joint_loss}) is non-convex, Algorithm \ref{al2} would converge to distinct results under different initializations. To reach a good local minimizer and speed up the convergence, we investigate several initialization schemes. Specifically, we provide an informative scheme that initializes the task membership $\mathbf{U}^{(0)}$ using the Dawid Skene (DS) estimator \citep{dawid1979maximum}, and initializes the worker group membership $\mathbf{V}^{(0)}$ based on the accuracy evaluated on the DS-estimation. In addition, the latent factors are initialized by multiple Gaussian components with different mean vectors, where the group memberships are given by $\mathbf{U}^{(0)}$ and $\mathbf{V}^{(0)}$. Finally, the initial rotation matrices $\mathbf{O}$ are set to be identity matrices. We also compare this informative scheme with random initialization in the simulation studies, which shows that the informative initialization converges within fewer iterations and achieves higher prediction accuracies. More detailed discussion about initialization schemes and the corresponding empirical studies can be found in Section \ref{sec: numerical_sensitivity}.
    
    One technical issue in our algorithm is that the number of worker subgroups $D$ is unknown and needs to be pre-specified which is critical to the proposed framework as it is highly related to the high-quality worker subgroup identification. With a small specified $D$, the performance of the high-quality worker subgroup tends to be undermined because some low-quality workers are likely to be mixed into the high-quality worker subgroup. On the other hand, with a large specified $D$, the workers from the same subgroup could be divided into multiple subgroups, and consequently lowers the power of the high-quality workers.
    
    In practice, the number of worker subgroups can be obtained from prior information, such as the level of workers' professionalism \citep{deng2014bayesian}. If prior information is not available, we can select the subgroup number by data-driven approaches, such as the jump statistics \citep{sugar2003finding}, the gap statistic \citep{tibshirani2001estimating}, and information criterion based method \citep{ma2017concave}. In this paper, we use a modified Bayesian information criterion \citep{wang2007tuning} to tune the number of worker subgroups $D$, 
    \begin{align}
        \label{BIC}
        BIC(D) = \log(\frac{\hat{\mathcal{L}}}{|\Omega|}) + (D + k - 1)\frac{\log|\Omega|}{|\Omega|},
    \end{align}
    where $|\Omega|$ denotes the cardinality of $\Omega$, and $\hat{\mathcal{L}} = \mathcal{L}(\hat{\mathbf{A}}, \hat{\mathbf{B}}, \hat{\mathbf{O}})$ is the log-likelihood given observed labeling matrix $\mathbf{R}$. In addition, we also discuss about the robustness of the proposed method under different numbers of work subgroups in Section \ref{sec: robust_worker_group}, which shows that a relatively small number of worker group is preferable.
    
\begin{figure}
        \centering
        \includegraphics[width=0.5\textwidth]{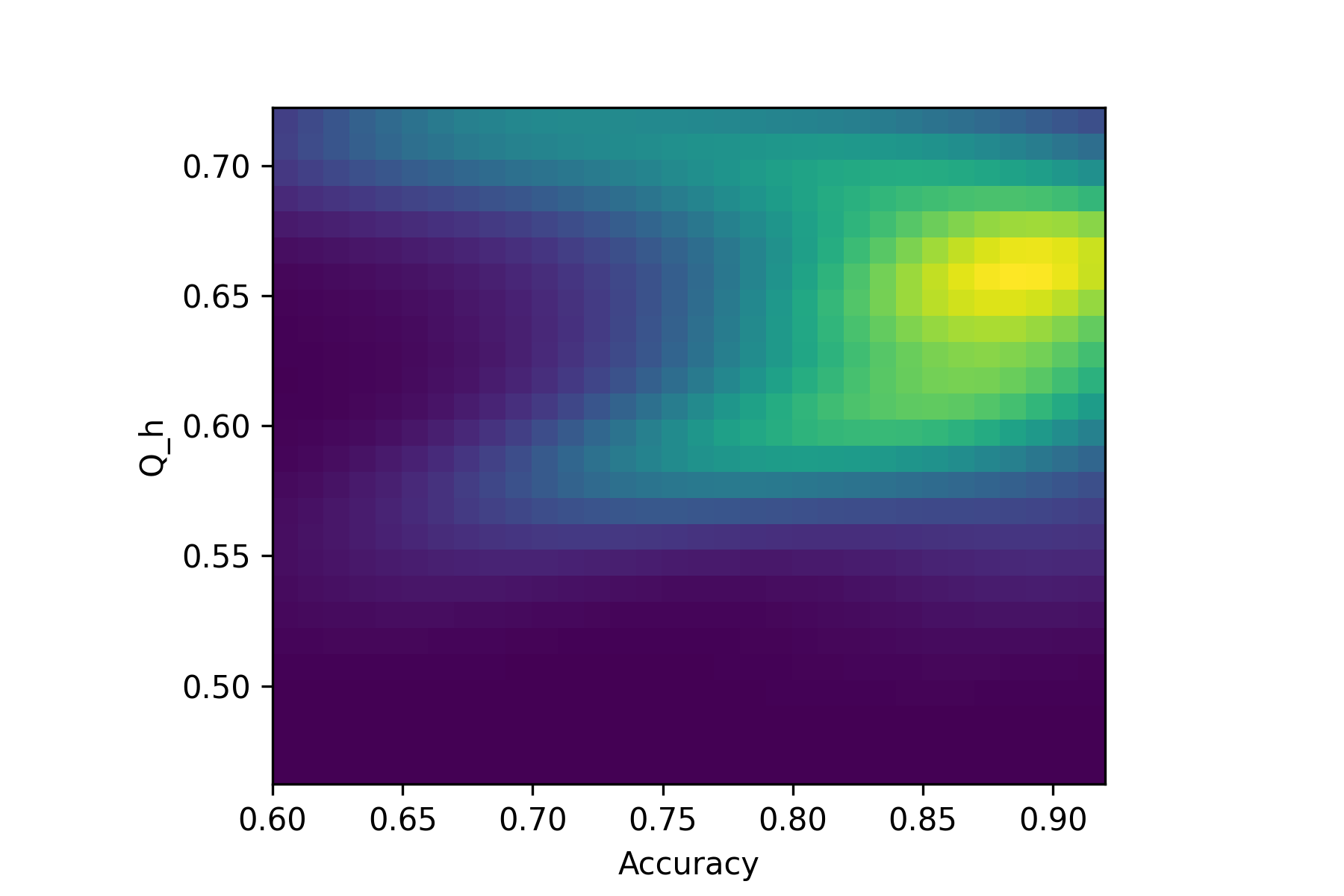}
        \caption{The kernel density plot of $Q_{h}$ versus accuracy based on the true labels. The experiment is based on 500 different combinations of $\lambda$ and $k$ with a fixed $D$ on the Bluebird data\citep{welinder2010multidimensional}.}
        \label{fig: criterion}
    \end{figure}

    Another challenge in crowdsourcing is lacking true label information, where the conventional cross validation \citep{stone1978cross} is not applicable. In the following, we propose a novel hyper-parameter tuning criterion for crowdsourcing without requiring any true label information. The proposed criterion utilizes Assumption \ref{assumption 2} in that some workers can provide more accurate labels than others. Therefore, we can identify the high-quality worker subgroups based on (\ref{eq: multi_criterion}). Consequently, we propose to select hyper-parameter $\lambda$ and $k$ in (\ref{eq: joint_loss}) and (\ref{eq: multi-categorical model}) via fitting labels based on those identified high-quality workers. Specifically, assuming that the set of indices of labels given by the high-quality workers is $\Omega_{h} = \{(i, j)\in\Omega| \hat{d}_{i} = V_{j}\}$. Then the criterion is defined as
    \begin{align}
        \label{Q_h}
        Q_{h}(\lambda, k)= \frac{\sum_{i,j\in \Omega_h}\mathbbm{1}({r_{ij} = \hat{r}_{ij}(\lambda, k)})}{|\Omega_h|},
    \end{align}
    where $\hat{r}_{ij}(\lambda, k)$ is the fitted labels via minimizing (\ref{eq: joint_loss}) given the penalization coefficient $\lambda$ and the dimension of latent factors $k$. The criterion $Q_{h}$ measures the goodness of estimated labels to labels given by the high-quality workers, and can be used to approximate the accuracy based on the true labels because crowd labels from the high-quality workers are more accurate and are good substitutes for the true labels. Figure \ref{fig: criterion} illustrates that $Q_{h}$ is highly correlated to the accuracy based on the true labels, suggesting that it performs well as a hyper-parameter tuning criterion without true label information.

    To apply the proposed criterion in (\ref{Q_h}), the high-quality workers need to be identified prior to tuning hyper-parameters $\lambda$ and $k$. We first select a proper number of worker subgroups through the modified BIC criterion (\ref{BIC}). Our empirical studies indicate that with a proper selected number of worker groups, the worker group memberships are not sensitive to $\lambda$ and $k$, in that the high-quality reference labels $\Omega_h$ is quite stable, although we can still perform a grid search to select an optimal hyper-parameter combination of $\lambda$ and $k$.

\section{Theory}
\label{sec: theory}
    In this section, we provide the theoretical guarantee of the proposed method in estimating latent factors and the group memberships. Specifically, we show that the proposed method achieves a faster convergence rate of estimation via imposing the multi-centroid grouping penalty compared with the $L_2$ penalty. We also show the consistency of the proposed method for recovering the true labels based on the assumptions on the high-quality workers. Here we provide the theoretical results for multicategory crowdsourcing only, and the results for the binary crowdsourcing can be derived as a special case.
    
    Consider the latent factor model (\ref{eq: multi-categorical model}) with the true parameter $\boldsymbol\theta^{*}$ following
    \begin{align}
        \label{model_true_para}
        \theta^{*}_{ijc} = \log\bigg\{\frac{p_{ij}^{(c)}}{p_{ij}^{(0)}}\bigg\} = {\mathbf{a}^{*}_{i}}'{\mathbf{O}_{c}^{(d)}}^{*}\mathbf{b}^{*}_{j}, \quad d = V^{*}_{j},
    \end{align}
    where $\mathbf{a}_{i}^{*} \in \mathbb{R}^{k}, \mathbf{b}_{j}^{*} \in \mathbb{R}^{k}$, ${\mathbf{O}_{c}^{(d)}}^{*} \in \mathbb{R}^{k\times k}$ and each ${\mathbf{O}_{c}^{(d)}}^{*}$ is an orthogonal matrix. The true group memberships of tasks and workers are denoted as $\mathbf{U}^{*}$ and $\mathbf{V}^{*}$, and their corresponding projection matrices are denoted as $\mathcal{U}^{*}$ and $\mathcal{V}^{*}$, respectively. 
    
    Let $\mathcal{PL}(\mathbf{A}, \mathbf{B}, \mathbf{O})$ be the penalized likelihood function:
    \begin{align}
        \mathcal{PL}(\mathbf{A}, \mathbf{B}, \mathbf{O}) = \mathcal{L}(\mathbf{A}, \mathbf{B}, \mathbf{O}) + \lambda_{|\Omega|}\mathcal{G}(\mathbf{A}, \mathbf{B}), \notag
    \end{align}
    where $\lambda_{|\Omega|}$ is the penalization parameter, $|\Omega|$ is the cardinality of $\Omega$, and $\mathcal{G}$ is the penalty function for the latent factors $\mathbf{A}$ and $\mathbf{B}$. In the subsequent analysis, we denote task latent factors as ${\boldsymbol\gamma}_{t} = (\mathbf{a}_1, \mathbf{a}_2, ..., \mathbf{a}_m)$ and worker latent factors ${\boldsymbol\gamma}_{w} = (\mathbf{O}_{0}^{(V_{1})}\mathbf{b}_{1}, \mathbf{O}_{1}^{(V_{1})}\mathbf{b}_{1}, ..., \mathbf{O}_{C-1}^{(V_{1})}\mathbf{b}_{1}, ..., \mathbf{O}_{C-1}^{(V_{n})}\mathbf{b}_{n})$. We consider two penalty functions and their resulting asymptotic properties: $\mathcal{G}_1(\mathbf{A, B}) = \lVert \mathbf{A}\rVert_F^2 + \lVert\mathbf{B}\rVert_F^2$ and $\mathcal{G}_2(\mathbf{A, B}) = \lVert \mathbf{A} - \mathcal{U}\mathbf{A} \rVert_F^2 + \lVert \mathbf{B} - \mathcal{V}\mathbf{B}\rVert_F^2$.
  
    We define the parameter space $\mathcal{S}$ of the latent factors $\boldsymbol\gamma = (\boldsymbol\gamma_{t}, \boldsymbol\gamma_{w})$ as follows. Let $\tau_1 > 0$ be a constant controlling the penalization on the latent factors. We assume that $\mathcal{G}(\boldsymbol\gamma) < \tau_1$, where $\mathcal{G}$ can be $\mathcal{G}_1$ or $\mathcal{G}_2$. In addition, since the probability of each label is unlikely to be 0, 1, or $C-1$ exactly, it is sensible to assume that $\lVert\boldsymbol\gamma\rVert_{\infty} \le \tau_2$. Then, we define the parameter space as:
    \begin{equation}
        \mathcal{S}(\tau_1, \tau_2) = \{\boldsymbol{\gamma}: \lVert \mathcal{G}(\boldsymbol{\gamma})\rVert \le \tau_1, \lVert\boldsymbol{\gamma} \rVert_{\infty}\le \tau_2 \}, \notag
    \end{equation}
    where the upper bound $\tau_1$ may depend on the sample size $m$ and $n$ and dimension of latent factors $k$.
    
    To measure the discrepancy between $\boldsymbol\gamma$ and $\tilde{\boldsymbol\gamma}$ in $\mathcal{S}(\tau_1, \tau_2)$, we introduce the hellinger distance as in \citep{shen1998method}:
    \begin{align}
        \mathcal{H}_{\mathcal{S}}(\boldsymbol{\gamma}, \tilde{\boldsymbol{\gamma}}) = \big\{\frac{1}{mn}\sum_{i=1}^m\sum_{j=1}^n[\int (f^{1/2}(r_{ij}; \boldsymbol\gamma) - f^{1/2}(r_{ij}; \tilde{\boldsymbol\gamma}))^2d\nu(r_{ij})]\big\}^{1/2}, \notag
    \end{align}
    where $f(r_{ij};\boldsymbol\gamma)$ is the likelihood function of $r_{ij}$ given the parameter $\boldsymbol\gamma$, and $\nu(\cdot)$ is a probability measure. The hellinger distance $\mathcal{H}_{\mathcal{S}}(\cdot, \cdot)$ can be shown as a valid distance function, and upper bounded by $\lVert \boldsymbol\gamma - \tilde{\boldsymbol\gamma}\rVert_F^2$.
    
    We consider the estimator $\hat{\boldsymbol\gamma}$ satisfying $\mathcal{PL}(\hat{\boldsymbol{\gamma}}\mid\mathbf{R}) \le \arg\min_{\boldsymbol\gamma\in\mathcal{S}(\tau_1, \tau_2)} \mathcal{PL}(\boldsymbol{\gamma}\mid\mathbf{R}) + \tau_{|\Omega|}$, where $\lim_{|\Omega|\rightarrow\infty}\tau_{|\Omega|} = 0$. In the proposed method, the loss function $\mathcal{PL}(\hat{\boldsymbol\gamma}|\mathbf{R})$ is bounded below by zero, so its infimum always exists and is finite. Theorem 1 provides the theoretical guarantee that the estimator $\hat{\boldsymbol\gamma}$ converges to the true parameter $\boldsymbol\gamma^{0}$ in probability under certain conditions.
    
    \begin{theorem}
     Suppose $\lambda_{|\Omega|} < \frac{1}{\tau_{1}(m + n)}\epsilon_{|\Omega|}^2$, there exists a positive constant $c_1$ such that 
    \begin{equation}
        \label{convergence}
        P(\mathcal{H}_{\mathcal{S}}(\hat{\boldsymbol{\gamma}}, \boldsymbol{\gamma}^0) \ge \epsilon_{|\Omega|}) \le 7\exp(-c_1|\Omega|\epsilon_{|\Omega|}^2), 
    \end{equation}
    with the best possible convergence rate of $\hat{\boldsymbol\gamma}$ is
    \begin{equation}
        \label{rate}
        \epsilon_{|\Omega|} \sim \frac{\sqrt{(m+n)k}}{|\Omega|^{1/2}}\Bigg[\log\bigg\{\frac{\tau_{1}|\Omega|}{\sqrt{mnk}(m+n)}\bigg\}\Bigg]^{1/2}.
    \end{equation}
    \end{theorem}
    Here, $\tau_1$ in (\ref{rate}) is determined by the choice of penalty function, which quantifies the volume of parameter space $\mathcal{S}(\tau_1, \tau_2)$. Considering the penalty function $\mathcal{G}_1(\cdot)$, $\tau_1 \sim O((m+Cn)k)$ and the penalty function $\mathcal{G}_2(\cdot)$, $\tau_1 \sim O((m_0 + n_0)k)$, where $m_0$ is the maximum number of tasks in any category and $n_0$ is the maximum number of workers in any subgroup, Corollary 1 provides the convergence rate under $\mathcal{G}_{1}$ and $\mathcal{G}_{2}$ and show that the proposed method with the multi-centroid grouping penalty $\mathcal{G}_2$ achieves a faster convergence rate than the $L_2$ penalty $\mathcal{G}_1$ without subgrouping.
    
    \begin{corollary}
    Under the assumptions in Theorem 1, the best convergence rate of (\ref{convergence}) with respect to $\mathcal{G}_1$ is 
    \begin{equation}
        \label{G1_rate}
        \epsilon_{|\Omega|} \sim \frac{\sqrt{(m+n)k}}{|\Omega|^{1/2}}\Bigg[\log\bigg\{\frac{\sqrt{k}|\Omega|}{\sqrt{mn}}\bigg\}\Bigg]^{1/2},
    \end{equation}
    and for the penalty function $\mathcal{G}_2$, the best convergence rate is
     \begin{equation}
        \label{G2_rate}
        \epsilon_{|\Omega|} \sim \frac{\sqrt{(m+n)k}}{|\Omega|^{1/2}}\Bigg[\log\bigg\{\frac{\sqrt{k}|\Omega|(m_0 + n_0)}{\sqrt{mn}(m+n)}\bigg\}\Bigg]^{1/2}.
    \end{equation}
    \end{corollary}
    
    \begin{remark}
    The above theorem and corollary are built upon the estimator $\hat{\boldsymbol\gamma}$ satisfying the condition $\mathcal{PL}(\hat{\boldsymbol{\gamma}}\mid\mathbf{R}) \le \arg\min_{\boldsymbol\gamma\in\mathcal{S}(\tau_1, \tau_2)} \mathcal{PL}(\boldsymbol{\gamma}\mid\mathbf{R}) + \tau_{|\Omega|}$, where the global minimizer of $\mathcal{PL}(\cdot\mid\mathbf{R})$ is not required, which is a relatively relaxed assumption for non-convex minimization.
    \end{remark}
    
    \begin{remark}
    The convergence in (\ref{convergence}) is built the the Hellinger distance $\mathcal{H}_{\mathcal{S}}(\cdot, \cdot)$ over the parameter space $\mathcal{S}$. The convergence property under the $L_2$ criterion is more restrictive than the convergence under the Hellinger distance. Specifically, under the Hellinger distance, the convergence of the estimator only depends on the size of the parameter space $\mathcal{S}$ and penalty coefficient $\lambda_{|\Omega|}$. In contrast, the convergence under the $L_2$ distance requires a stronger local property in that $Var(L(\boldsymbol\gamma) - L(\boldsymbol\gamma^{0}))$ is bounded over the parameter space.
    \end{remark}
    
    \begin{remark}
    The improvement of $\mathcal{G}_2(\cdot)$ over $\mathcal{G}_1(\cdot)$ regarding the convergence rate is illustrated by the $\frac{m_0 + n_0}{m + n}$ in the logarithm term in (\ref{G2_rate}). Under a uniform bound $\tau_{2}$ on the magnitude of latent factors, the parameter space constrained by $\mathcal{G}_2(\cdot)$ is in the same order of the parameter space containing the largest subgroup when the underlying subgroup structure exists. In contrast, the penalty $\mathcal{G}_1(\cdot)$ is a special case with only one group for all latent factors. Therefore the parameter space constrained by $\mathcal{G}_2(\cdot)$ is smaller than the parameter space constrained by $\mathcal{G}_1(\cdot)$. Since the Hellinger distance has an advantage that the convergence rate of $\hat{\boldsymbol\gamma}$ only depends on the size of the parameter space and the penalization coefficient $\lambda_{|\Omega|}$ \citep{shen1998method}, the convergence rate under $\mathcal{G}_2(\cdot)$ is faster than under $\mathcal{G}_1(\cdot)$.
    \end{remark}
    \begin{remark}
    In order to obtain a consistent estimation of the model parameters and the tasks' labels, we require that the $\epsilon_{|\Omega|}$ goes to zero as the number of tasks $m$ and the number of workers $n$ approach infinity. Therefore, we can derive the order of $|\Omega|$ from the best rate of convergence from Corollary 1. Specifically, if we use the $L_2$ penalty for latent factors estimation, then $|\Omega| \ge \mathcal{O}((m+n)k\log\{(m+n)k\})$ guarantees the convergence of $\hat{\boldsymbol\gamma}$ and the consistent estimation of group memberships of tasks and workers; and if we use the proposed multi-centroid grouping penalty, then $|\Omega| \ge \mathcal{O}((m+n)k\log\{(m_0+n_0)k\})$ (where $m_0$ and $n_0$ are the maximum cardinality of tasks and workers groups, respectively) guarantees the same theoretical properties of $\hat{\boldsymbol\gamma}$ and group memberships. It is noteworthy that utilizing the MCGP requires fewer observed labels compared with the $L_2$ penalty, as the proposed MCGP is able to borrow information from the same group of individuals.
    \end{remark}
    
    Next, we establish the consistency of the subgroup memberships. Let $M_{k}^{*} = \{i: \mathbf{U}^{*}_{i}=k\}$ be the $k$th true task category, and $N_{k}^{*} = \{j: \mathbf{V}_{j}^{*}=k\}$ be the $k$th true worker subgroup. Similarly, we define $\hat{M}_{k}$ and $\hat{N}_{k}$ as the $k$th estimated task category and worker subgroup, respectively. Noting that the subgroup memberships are invariant for label permutations, we define the permutation operation as $\pi(\cdot)$ and the estimation error of subgroup memberships as
    \begin{align}
        err_{\text{task}} = \min_{\pi(\cdot)}\frac{1}{m}\sum_{k=0}^{C-1}|M_{\pi(k)}^{*}\backslash\hat{M}_{k}|, \quad err_{\text{worker}} = \min_{\pi(\cdot)}\frac{1}{n}\sum_{k=1}^{D}|N_{\pi(k)}^{*}\backslash\hat{N}_{k}|, \notag
    \end{align}
    where $|\cdot|$ is the cardinality of a set, and $A\backslash B$ denotes the set $\{i: i\in A \text{ and } i\notin B\}$.
    
    \begin{theorem}
    Suppose assumptions in Theorem 1 holds, and there exists constant $\eta_1> 0$ such that $\frac{1}{m}\lVert \mathbf{A}^{*} - \mathcal{U}^{*}\mathbf{A}^{*}\rVert_{2\rightarrow\infty} \le \eta_{1}$ and $\frac{1}{n}\lVert \mathbf{B}^{*} - \mathcal{V}^{*}\mathbf{B}^{*}\rVert_{2\rightarrow\infty} \le \eta_{1}$ where $\lVert \mathbf{A} \rVert_{2\rightarrow\infty} = \sup_{\lVert\mathbf{x}\rVert=1}\lVert\mathbf{Ax}\rVert_{\infty}$ is the so-called two-to-infinity norm \citep{cape2019two}. Then the estimated subgroup memberships $\hat{\mathbf{U}}$ and $\hat{\mathbf{V}}$ are consistent:
    \begin{align}
        err_{task} \rightarrow_{P} 0, \quad err_{worker} \rightarrow_{p} 0. \notag
    \end{align}
    \end{theorem}
    
    \begin{remark}
    The constant $\eta_1$ in Theorem 2 controls the distance between individual latent factors and its corresponding group centroids, which measures the concentration of the task and worker groups. Since the latent factor modeling (\ref{eq: multi-categorical model}) aims to capture the heterogeneity within tasks and workers, a positive $\eta_1$ is required for $\mathbf{A}$ and $\mathbf{B}$, which is quite different from the network embedding in \citep{zhang2021directed}. In their theoretical analysis, they follow the stochastic block model \citep{rohe2016co} where the same group nodes share identical latent factors, and which is equivalent to $\eta_1 = 0$ in our model.
    \end{remark}

    \begin{corollary}
    Suppose assumptions in Theorem 1 and Theorem 2 holds, and Assumption \ref{assumption 2} (high-quality worker subgroup identification assumption) holds, the predicted labels $\hat{\mathbf{Z}}$ converges to $\mathbf{Z}$ in probability.
    \end{corollary}
    
    Theorem 2 ensures that the proposed estimators of subgroup memberships are consistent, indicating that tasks from different categories can be clustered into different subgroups asymptotically. Similarly, the workers' subgroup memberships can also be recovered asymptotically. In summary, the correspondence between the true labels and task memberships can be recovered asymptotically based on Assumption \ref{assumption 2} through the identification of the high-quality worker subgroup. Therefore, the predicted labels are consistent estimators of the true labels. We provide the proofs of theorems and corollaries in the supplementary materials. 
    
\section{Simulation Studies}
\label{sec:num_exp}

    In this section, we conduct simulations for multicategory crowdsourcing to illustrate the performance of the proposed method. Particularly, we compare our method with three existing methods including the majority voting (MV), the Dawid-Skene estimator initialized by majority voting (DS-MV) \citep{dawid1979maximum}, the Dawid-Skene estimator initialized by spectral method (DS-Spectral) \citep{zhang2016spectral}. In the supplementary materials, we consider more simulation settings to investigate the performance of the proposed method in binary crowdsourcing problems, and compare with two more methods which are specifically designed for binary crowdsourcing.

\subsection{Study 1: Multicategory Crowdsourcing}
\label{sim: multicategory}
    In this study, we consider the multicategory crowdsourcing with three task categories ($C = 3$) and three worker groups ($D = 3$). In the multicategory case, high-quality worker group for different task category might not be the same.
    
    \begin{table}
        \caption{Expertise assignment and latent factor centroid of each worker group on task categories for Study 2. "High" represents the high probability of labeling the specific task category correctly, "low" represents the high probability of labeling the specific task category incorrectly, and "average" represents expertise level that are better than "low" but worse than "high".}
        \begin{center}
        \small
        \begin{tabular}{cccc}
        \hline
        \hline
                            & Worker Group 1     & Worker Group 2      & Worker Group 3 \\
        \hline
        label-0 task category    &  high              &   low               &  average       \\
        $(2, 0, 0)'$             & $(2, 0, 0)'$       & $(0, 1, 1)'$        & $(1, 1, 0)'$      \\
        \hline
        label-1 task category     &  average           &   high              &  average       \\
        $(0, 2, 0)'$             & $(1, 1, 1)'$       &$(0, 2, 0)'$         &$(1, 1, 0)'$   \\
        \hline
        label-2 task category    &  low               &   low               &  high          \\
        $(0, 0, 2)'$             & $(1, 1, 0)'$      & $(1, 1, 0)'$         & $(0, 0, 2)'$ \\
        \hline
        \hline
        \end{tabular}
        \end{center}
        \label{tab: multi-sim}
    \end{table}

    Let the number of tasks $m = 150$ with 50 tasks in each category, and the number of workers $n = 150$ with 50 workers in each group. The latent factors of tasks $\mathbf{a}_i$'s are generated from a multivariate normal distribution $\mathcal{N}(\boldsymbol\alpha_c, \sigma^2\mathbf{I}_3)$, where $\boldsymbol\alpha_0 = (2, 0, 0)'$ for label-1 tasks, $\boldsymbol\alpha_1 = (0, 2, 0)'$ for label-2 tasks and $\boldsymbol\alpha_2 = (0, 0, 2)'$ for label-3 tasks. In addition, the variance $\sigma^2 \in \{0.5, 2\}$ controls the task heterogeneity, where $\sigma^2 = 0.5$ and $\sigma^2 = 2$ indicate the homogeneous and heterogeneous scenarios, respectively. The latent factors of workers $\mathbf{b}_{j, c}$'s are generated from a multivariate normal distribution $\mathcal{N}(\boldsymbol\beta_{c, d}, \mathbf{I}_2)$ where $\boldsymbol\beta_{c, d}$ is centroid of the $d$th worker group on label-$c$ tasks. The centroid $\boldsymbol\beta_{c, d}$ is associated with the expertise of each worker group corresponding to each task category,  and a high concordance with task centroid indicates a high level of expertise. Various settings of workers' expertise on each task category and latent factor centroids of workers are provided in Table \ref{tab: multi-sim}. The crowd labels $r_{ij}$'s are  generated from multinomial distribution with parameters $(\tilde{p}_{ij,1}, \tilde{p}_{ij,2}, \tilde{p}_{ij, 3})$ where $\tilde{p}_{ij, c} = \frac{exp(\mathbf{a}_i'\mathbf{b}_{j,c})}{\sum_{c=1}^3 exp(\mathbf{a}_i'\mathbf{b}_{j,c})}$ represents the probability that $r_{ij}$ equals to $c$. Each crowd label is missing at random with probability of $0.7$.

    We compare the proposed method with MV, DS-MV and DS-Spectral. For the proposed method, we tune latent factor dimension $k$ ranging from 2 to 5, the number of worker group $D$ ranging from 2 to 6, and penalty coefficient $\lambda$ ranging from 0.001 to 1. The label prediction accuracies and standard errors are reported in Table \ref{sim2}. It is clear that the proposed method outperforms other methods in both homogeneous and heterogeneous scenarios. In the homogeneous scenario, the proposed method achieves 4.38$\%$ (DS-Spectral) to 43.4$\%$ (MV) improvement on accuracy; and in the heterogeneous scenario, the proposed method improves other methods from 6.13$\%$ (DS-MV) to 24.4$\%$ (MV) on accuracy.
    
    \begin{table}
        \caption{Label prediction accuracy (standard error) of the proposed method compared with three existing methods. The simulation results are based on 200 independent experiments.}
        \begin{center}
        \small
        \begin{tabulary}{.9\textwidth}{CCCCC}
        \hline
        \hline
        	$\sigma^2$      &The Proposed Method	&MV		&DS-MV			    &DS-Spectral\\
        \hline
        	0.5             &\textbf{0.928(0.050)}			&0.647(0.089)		&0.886(0.040)	&0.889(0.039)   \\
        	2	            &\textbf{0.745(0.030)}		&0.599(0.129)		&0.702(0.045)	&0.701(0.051)	\\
        \hline
        \hline
        \end{tabulary}
        \label{sim2}
        \end{center}
    \end{table}

    The improvement of the proposed method is due to the following reasons. First, the proposed method is able to identify the high-quality worker group for each task category with a high accuracy, and relies on the high-quality workers heavily for assigning labels. In particular, the proposed method achieves 75.5$\%$ and 73.2$\%$ accuracy on worker group classification under the homogeneous and heterogeneous scenarios. In contrast, the Dawid-Skene estimator relies on majority workers to assign labels regardless of their qualities, resulting in low prediction accuracy especially for tasks where majority worker subgroups have low-quality workers. Second, in the heterogeneous scenario, latent factor modeling of the proposed method incorporates the heterogeneity among tasks within the same category, while majority voting and confusion matrix-based methods do not take this into consideration.
    
    \subsection{Study 2: Robustness against number of worker subgroups}
    \label{sec: robust_worker_group}
    
    In this section, we study the robustness of the proposed method against mis-specified number of worker groups. In general, it is crucial to select a proper number of clusters to derive a interpretable or sensible clustering result. In the proposed framework, the number of task groups is pre-specified as the number of task classes, while the number of worker groups is to be determined in our algorithm. In the following, we investigate the robustness against the number of worker groups under several scenarios for multicategory tasks.
    
    We consider four scenarios of 3-category tasks other than the settings in Section \ref{sim: multicategory}. The detailed model settings are shown in Table \ref{tab: multicategory_simulation_generation}. Similar to the above setting, we assume there are high-quality, mid-quality and low-quality worker subgroups. We simulate 150 tasks with 50 tasks within each class, and 300 workers evenly distributed in each group. The simulations are repeated 200 times for each scenario.

\begin{table}[t]
    \small{
    \centering
    \begin{tabular}{p{20mm}p{25mm}p{40mm}p{30mm}p{50mm}}
    \hline
    Scenario & \#task groups & task group centroids & \#worker groups & worker group centroids \\
    \hline
    \multirow{4}{25mm}{1} & \multirow{4}{25mm}{3} & \multirow{4}{35mm}{label-1: $(2, 0, 0)$ label-2: $(0, 2, 0)$ label-3: $(0, 0, 2)$} & \multirow{4}{30mm}{2} & high-quality: $(2, 0, 0), (0, 2, 0), (0, 0, 2)$ \\
    & &  & & mid-quality: $(2, 0, 0), (0, 2, 1), (0, 0, 2)$\\
    \hline
    \multirow{4}{25mm}{2} & \multirow{4}{25mm}{3} & \multirow{4}{35mm}{label-1: $(2, 0, 0)$ label-2: $(0, 2, 0)$ label-3: $(0, 0, 2)$} & \multirow{4}{30mm}{2} & high-quality: $(2, 0, 0), (0, 2, 0), (0, 0, 2)$ \\
    & &  & & low-quality: $(1, 2, 0), (0, 2, 1), (1, 1, 1)$\\
    \hline
    \multirow{6}{25mm}{3} & \multirow{6}{25mm}{3} & \multirow{6}{35mm}{label-1: $(2, 0, 0)$ label-2: $(0, 2, 0)$ label-3: $(0, 0, 2)$} & \multirow{6}{30mm}{3} & high-quality: $(2, 0, 0), (0, 2, 0), (0, 0, 2)$ \\
    & &  & & mid-quality: $(2, 0, 0), (0, 2, 1), (0, 0, 2)$\\
    & &  & & low-quality: $(1, 2, 0), (0, 2, 1), (1, 1, 1)$\\
    \hline
    \multirow{6}{25mm}{4} & \multirow{6}{25mm}{3} & \multirow{6}{35mm}{label-1: $(2, 0, 0)$ label-2: $(0, 2, 0)$ label-3: $(0, 0, 2)$} & \multirow{6}{30mm}{3} & high-quality: $(2, 0, 0), (0, 2, 0), (0, 0, 2)$ \\
    & &  & & low-quality: $(2, 0, 0), (1, 0, 1), (0, 2, 0)$\\
    & &  & & low-quality: $(1, 2, 0), (0, 2, 1), (1, 1, 1)$\\
    \hline
    \end{tabular}
    }
    \caption{Model Settings for multicategory tasks}
    \label{tab: multicategory_simulation_generation}
\end{table}

\begin{table}[t]
\centering
    \small{
    \begin{tabular}{p{25mm}p{25mm}p{25mm}p{25mm}p{25mm}}
    \hline
     & \multicolumn{4}{c}{Specified number of worker groups} \\
    \hline
    Scenario & 2 & 3 & 4 & 5 \\
    \hline
    1 & 0.884(0.194) & 0.888(0.187) & 0.887(0.186) & \textbf{0.888(0.028)}\\
    2 & \textbf{0.893(0.116)} & 0.892(0.181) & 0.888(0.186) & 0.886(0.017)\\
    3 & 0.941(0.023) & 0.946(0.028) & \textbf{0.956(0.021)} & 0.953(0.022)\\
    4 & 0.824(0.239) & 0.945(0.027) & \textbf{0.945(0.026)} & 0.944(0.026)\\
    \hline
    \end{tabular}
    }
    \caption{Simulation results: robustness against the number of worker groups, label prediction accuracy of multi-categorical tasks}
    \label{tab: multicategory_simulation_result}
\end{table}

Simulation results are shown in Table \ref{tab: multicategory_simulation_result}, which indicates that the proposed method is quite robust against the mis-specification of the number of worker groups. For all four scenarios, prediction accuracies under different numbers of worker groups are pretty stable, and a higher number of worker groups even achieves a much lower standard deviation. With a smaller number of worker groups, some low-quality workers would be grouped into the high-quality worker group, so that the labeling certainty of the whole group is decreased. Therefore, we recommend using a relatively large number of worker groups in real applications.

\subsection{Study 3: Numerical sensitivity to initialization}
\label{sec: numerical_sensitivity}

Due to the non-convexity of the optimization problem (\ref{eq: joint_loss}), in general, it is infeasible to obtain the global minimizer numerically and the solutions obtained from the algorithm depend on the initialization. Therefore, it is essential to investigate the sensitivity of the proposed method to different initializations and propose an efficient initialization scheme to obtain great empirical results. 
    
In this subsection, we investigate several initialization schemes for latent factors and subgroup memberships and compare their empirical results. Specifically, we employ three schemes to initialize group memberships and latent factors in the proposed model: random, majority voting and Dawid-Skene estimation \citep{dawid1979maximum}, where the last two schemes are informative schemes to provide a warm initial estimation from the given data. Under each scheme, we first initialize the task memberships $\mathbf{U}^{(0)}$, and the three schemes adopt different task membership initializations: randomly generated membership, majority voting or Dawid-Skene estimation \citep{dawid1979maximum}. Next, we initialize the worker memberships according to the labeling accuracy for the three initialized task memberships. Specifically, the $j$th worker's group membership is specified as follows:
\begin{align}
    \mathbf{V}_{j}^{(0)} = \begin{cases}
    1 & acc(\mathbf{R}_{\cdot j}, \mathbf{U}^{(0)}) \le q_{\frac{1}{D}} \\
    2 &  q_{\frac{1}{D}} < acc(\mathbf{R}_{\cdot j}, \mathbf{U}^{(0)}) \le q_{\frac{2}{D}} \\
    ... &  \\
    D & q_{\frac{D-1}{D}} < acc(\mathbf{R}_{\cdot j}, \mathbf{U}^{(0)}) \le 1 \\
    \end{cases}, \notag
\end{align}
where $acc(\mathbf{R}_{\cdot j}, \mathbf{U}^{(0)}) = \frac{1}{|\mathbf{R}_{\cdot j}|}\sum_{i}I(\mathbf{R}_{ij} = \mathbf{U}^{(0)}_i)$ is the accuracy of the $j$th worker evaluated at the initialized task group membership $\mathbf{U}^{(0)}$. The thresholds $q_{\frac{1}{D}}, q_{\frac{2}{D}}, ..., q_{\frac{D-1}{D}}$ are the $\frac{1}{D}, \frac{2}{D}, ..., \frac{D-1}{D}$ quantiles of the empirical distribution of accuracies over all workers.
    
\begin{table}
    \centering
    \caption{Multicategory crowdsourcing simulation results of the proposed method under different initialization schemes. Random scheme: random initialization of task/worker memberships and latent factors; MV scheme: majority voting-based initialization of task memberships and associated worker memberships and latent factors; DS scheme: DS estimation-based initialization of task memberships and associated worker memberships and latent factors. Here, stability is the average of the mean and standard errors of prediction accuracy within each experiment.}
    {\small
    \begin{tabular}{c|c|c|c|c}
    \hline
    \hline
        Initialization & Orth mat initialization & Iterations to converge & Accuracy & Stability \\
    \hline
        Random scheme & Random & 14.82(6.98) & 0.595(0.097) & 0.354(0.115) \\
        MV scheme     & Identity & 12.79(6.41) & 0.697(0.059) & 0.575(0.088) \\
        DS scheme     & Identity & \textbf{11.46(5.11)} & \textbf{0.745(0.030)} & \textbf{0.669(0.070)} \\
    \hline
    \end{tabular}
    }
    \label{tab: multi_sens_init}
\end{table}
    
For the initialization of the latent factors of tasks and workers, we adopt different approaches in random and informative (MV or DS \citep{dawid1979maximum}) schemes. In the random scheme, we initialize latent factors following the random standard normal distribution in $\mathbb{R}^{k}$ regardless of the group memberships. In the informative schemes, we initialize latent factors using different normal distribution in $\mathbb{R}^{k}$ where each group specified by $\mathbf{U}^{(0)}$ or $\mathbf{V}^{(0)}$ has its own random unit-length means, and the covariance matrices are specified as identity matrices. In addition, we compare two schemes to initialize the orthogonal matrices in multicategory crowdsourcing: (1) identity matrices for all $\mathbf{O}^{(d)}_{c}$'s; (2) random orthogonal matrices except for the fixed reference group $\mathbf{O}^{(d)}_{0}$'s.
    
In the following, we report the simulation results based on the above initialization schemes. we conduct simulations based on the heterogeneous setting ($\sigma^2 = 2$) as in Section \ref{sim: multicategory}. We run 100 independent experiments and generate 10 different initial values of group memberships and latent factors within each experiment. We report the performance criteria such as iterations, accuracy and stability in Table \ref{tab: multi_sens_init}. The informative schemes converge faster than the random scheme, and achieve better prediction accuracy and stability.
    
In summary, our numerical experiments show an informative initialization scheme is essential to guarantee a faster convergence and empirical performance. Therefore, we recommend using informative schemes in practice to achieve faster and more accurate predictions.

\section{Real Data Application}

    In this section, we apply the proposed method to five real data sets and compare our performance with several competitive methods. Among the five datasets, Bluebird \citep{welinder2010multidimensional}, Glaucoma \citep{mitry2015crowdsourcing} and Breaset cancer \citep{street1993nuclear} are binary tasks, whereas Iris \citep{fisher1936use} and Dog \citep{deng2009imagenet} are multicategory tasks. Bluebird, Glaucoma and Dog datasets are labelled by human workers from online platform, while Breast cancer and Iris datasets are labelled by multiple machine learning algorithms. The summary statistics for the five datasets are provided in Table \ref{tab: data-stat}. 
    
    \begin{table}
        \begin{small}
        \caption{Summary statistics of five real datasets, including the number of categories, the number of tasks, the number of workers, the number of crowd labels and the missing rate of crowd label matrix.}
        \begin{center}
        \scriptsize
        \begin{tabular}{cccccc}
        \hline
        \hline
            Dataset         &\#Categories        &\#Tasks        &\#Workers       &\#Crowd labels    &Missing rate  \\
            BlueBird        &2               &108           &39               &4,212            & $0.0\%$   \\            
            Glaucoma        &2               &127           &78               &2,540            & $74.4\%$\\           
            Breast Cancer   &2               &500           &24               &12,000           & $0.0\%$\\   
            \hline
            Iris            &3               &100           &24               &2,000            & $16.7\%$           \\   
            Dog             &4               &807           &52               &7,354            & $82.5\%$             \\  
        \hline
        \hline
        \end{tabular}
        \label{tab: data-stat}
        \end{center}
        \end{small}
    \end{table}

    The bluebird dataset includes 108 images and workers are assigned to distinguish whether a bird is visible in the images. The Glaucoma dataset includes normal and glaucomatous disc images with disease status pre-determined by experts. Workers are trained with a simple tutorial and assigned to distinguish the disease status in the disc images. The true labels for each image are not available, however, we can predict true labels via summary statistics in \citep{mitry2015crowdsourcing}. Breast cancer dataset includes hand-crafted features from digitized images of a fine needle aspirate (FNA) of a breast mass, and corresponding cancer status, such as malignant or benign. The dataset consists of 569 observations, where the first 69 observations are used for training 24 different binary classifiers listed in the Appendix. These 24 classifiers are treated as crowd workers, and the predicted labels of the remaining 500 observations are treated as crowd labels.

    Iris dataset contains three types of iris and their feature measurements. We use the first 50 observations for training 24 different classifiers, and predict the remaining 100 observations with the trained classifiers for crowd labels. Dog dataset consists of four breeds of dogs, and workers are asked to label each dog image as norfolk terrier, norwich terrier, irish wolfhound, or scottish deerhound.
    
    In addition to the methods specified in Section \ref{sec:num_exp}, we also compare our method with two more competing methods: Mixed Strategies Crowdsourcing (MiSC \citep{ko2019misc}) and Community-based Bayesian Classifier Combination (CBCC, \citep{venanzi2014community}), where the MiSC method decomposes an augmented tensor and the CBCC employs subgroup structures for the workers' confusion matrices. The comparisons of these methods are summarized in Table \ref{tab: real}, showing that the proposed method consistently outperforms other methods on label prediction accuracy. In particular, the improvement of the proposed method over other methods ranges from 3.1$\%$ to 29.5$\%$ on Bluebird dataset, 1.6$\%$ to 76.9$\%$ on Glaucoma dataset, and 2.36$\%$ to 4.95$\%$ on the Breast cancer dataset. For the multicategory dataset, the proposed method achieves 3.2$\%$ to 7.07\% improvement on Iris dataset and 0.0$\%$ to 5.22$\%$ improvement on Dog dataset.
    
    \begin{table}
        \begin{small}
        \caption{True label prediction accuracy of the proposed method compared with existing methods in real data. "MV" is the majority voting, "DS-MV" is the majority voting initialized dawid-skene estimator, "DS-Spectral" is the spectral initialized dawid-skene estimator, "MiSC" is the Mixed Strategies Crowdsourcing, "CBCC" is the Community-based Bayesian Classifier Combination, "SVD" is the singular value decomposition, "GLAD" is the generative label ability and difficulty model.}
        \begin{center}
        \scriptsize
        \begin{tabular}{cccccccccc}
        \hline
        \hline
            Dataset     &Category        &The Proposed Method    &MV         &DS-MV      &DS-Spectral    & MiSC  & CBCC    &SVD       &GLAD  \\
            BlueBird    &2               &\textbf{0.935}         &0.759      &0.889      &0.899   &0.907 &0.889       &0.722     &0.789 \\
            Glaucoma    &2               &\textbf{0.913}         &0.516      &0.890      &0.787  &0.899  &0.874    &0.567     &0.726 \\
            Breast Cancer   &2           &\textbf{0.868}         &0.827      &0.848      &0.848     &0.844 &0.838         &0.840     &0.836 \\
            \hline
            Iris        &3               &\textbf{0.960}         &0.890      &0.920      &0.920   & 0.930    &0.920      &-  &-\\
            Dog         &4               &\textbf{0.846}         &0.804      &0.833      &0.831    &\textbf{0.846}    &0.814     &-  &-\\
        \hline
        \hline
        \end{tabular}
        \label{tab: real}
        \end{center}
        \end{small}
    \end{table}

\begin{figure}
        \begin{subfigure}{0.5\textwidth}
        \includegraphics[width=0.9\linewidth, height=5cm]{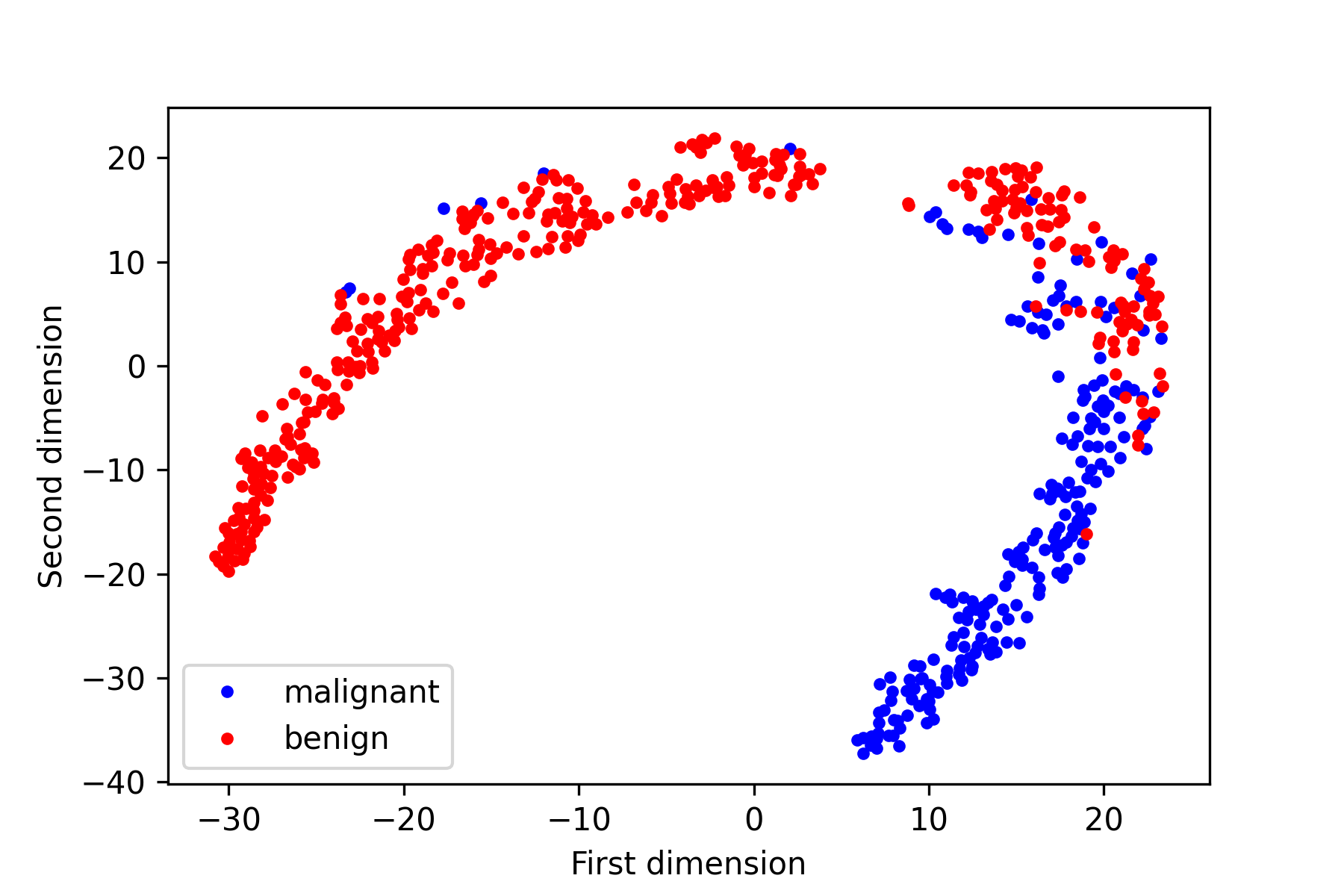} 
        \end{subfigure}
        \begin{subfigure}{0.5\textwidth}
        \includegraphics[width=0.9\linewidth, height=5cm]{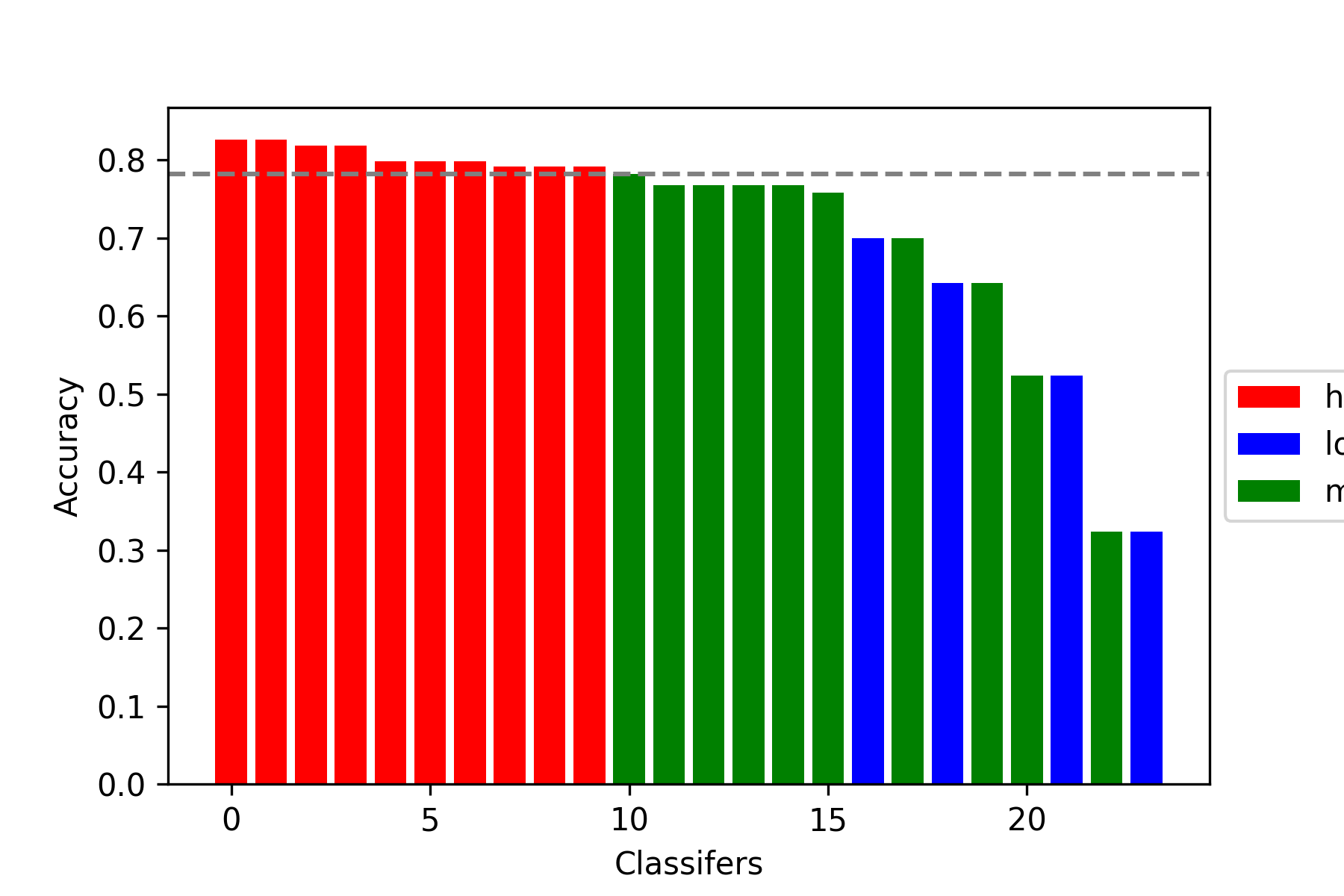}
        \end{subfigure}
        \caption{Breast cancer dataset. Left panel: data points are transformed by t-SNE \citep{maaten2008visualizing}; Right panel: classifier accuracy.}
        \label{fig: breast_cancer}
    \end{figure}

    The proposed method outperforms other methods due to the following reasons. First, tasks in the same category could be heterogeneous. For example, in Figure \ref{fig: breast_cancer}, the left panel provides visualization of first two latent factors from dimension reduction via t-stochastic neighbor embedding (t-SNE) \citep{maaten2008visualizing} for Breast cancer data, where latent features of either benign or malignant categories are scattered in the latent space. The proposed method is able to incorporate heterogeneity using latent factor modeling. Second, the expertise levels of workers could be quite different. The right panel of Figure \ref{fig: breast_cancer} provides the accuracy of each individual trained classifier, indicating that certain classifiers perform better than the others, and the proposed method can identity high-quality workers to improve the label prediction accuracy.
    
\section{Discussion}

    In this paper, we propose a novel crowdsourcing framework for multicategory classification. The proposed method is able to incorporate heterogeneity among tasks and workers via the latent factor modeling. We also utilize the subgroup structures in the latent space through the multi-centroid grouping penalty and group-specific rotation matrices. Therefore, tasks can be clustered based on their categories, and workers can be differentiated based on their opinions on different tasks. Compared to existing crowdsourcing approaches, the proposed method identifies high-quality workers with less restrictive assumptions and provides a labeling specification procedure that is more robust against noisy crowd labels.

    In theory, we establish the estimation consistency of model parameters under hellinger distance. In contrast to the $L_2$ penalty, the proposed method with the multi-centroid grouping penalty can achieve a faster convergence. In addition, we show the consistency of labeling prediction asymptotically. For a finite sample problem, our simulation studies illustrate that the proposed method improves label prediction accuracy by incorporating heterogeneity and subgroup structures among tasks and workers. The improvement is more significant when the proportion of low-quality workers is relative high to high-quality workers, which benefits from the robust label specification procedure.

    Several extensions are worth further exploring in the future. In this paper, we only consider the linear concordance between task and worker latent factors. However, this can be generalized via nonlinear matrix completion techniques \citep{fan2018matrix} for modeling more complex task-worker relations in practice. In addition, on platforms such as Mturk, high-quality workers are usually encouraged to label more tasks with incentives, and the crowd labels are not missing at random. Therefore, the new method should be able to take the informative missing mechanisms into consideration. In addition, Assumption \ref{assumption 2} of high-quality workers is not statistically testable, which is likewise the commonly adopted majority voting. Therefore, it is also interesting to investigate a valid procedure to verify this assumption.

\section*{Acknowledge}

The authors thank the Editor, Associate Editor and the anonymous reviewers for their insightful suggestions and helpful feedback which improved the paper significantly. The authors declare no financial or non-financial interest that has arisen from the direct applications of this research. This work is supported by NSF Grants DMS 2210640, DMS 1952406, HK RGC Grants GRF-11304520, GRF-11301521, GRF-11311022, and CUHK Startup Grant 4937091.

\section*{Supplementary Materials}

The supplementary materials provide the methodology for binary crowdsourcing, simulation results for binary crowdsourcing, and proofs of theorems and corollaries.

\newpage
{\footnotesize\bibliography{reference}}

\end{document}